\documentclass[aps,prl,twocolumn,showpacs,10pt,superscriptaddress,preprintnumbers,nofootinbib]{revtex4-1}
\usepackage{epsfig,amssymb,amsmath,psfrag,epstopdf,color}
\pdfoutput=1
\usepackage{graphicx}
\usepackage{subfig}
\usepackage[margin=7pt,justification=centerlast]{caption}

\usepackage{paralist}
\usepackage{hyperref}

\usepackage{slashed} 

\allowdisplaybreaks

\def\beq{\begin{equation}}
\def\eeq{\end{equation}}
\def\bsp#1\esp{\begin{split}#1\end{split}}
\newcommand{\be}{\begin{equation}}
\newcommand{\ee}{\end{equation}}
\newcommand{\bea}{\begin{eqnarray}}
\newcommand{\eea}{\end{eqnarray}}

\def\Fig#1{Fig.~{\ref{#1}}}

\def\to{\rightarrow}

\def\bea#1\eea{\begin{align}#1\end{align}}

\def\ksl{\not{\hbox{\kern-2.3pt $k$}}}

\def\spa#1.#2{\left\langle#1\,#2\right\rangle}
\def\spb#1.#2{\left[#1\,#2\right]}
\def\lor#1.#2{\left(#1\,#2\right)}
\def\sand#1.#2.#3{%
\left\langle\smash{#1}{\vphantom1}^{-}\right|{#2}%
\left|\smash{#3}{\vphantom1}^{-}\right\rangle}

\newcommand{\fd}[2]{\parbox{#1}{\includegraphics[width=#1]{#2}}}

\newcommand{\nn}{\nonumber}

\begin{document}

\preprint{MIT-CTP 5474}

\title{Beautiful and Charming Energy Correlators}

\author{Evan Craft}
\email{evan.craft@yale.edu}
\affiliation{Department of Physics, Yale University, New Haven, CT 06511}

\author{Kyle Lee}
\email{kylel@mit.edu}
\affiliation{Nuclear Science Division, Lawrence Berkeley National Laboratory, Berkeley, CA 94720}
\affiliation{Center for Theoretical Physics, Massachusetts Institute of Technology, Cambridge, MA 02139, USA}

\author{Bianka Me\c caj}
\email{bianka.mecaj@yale.edu}
\affiliation{Department of Physics, Yale University, New Haven, CT 06511}

\author{Ian Moult}
\email{ian.moult@yale.edu}
\affiliation{Department of Physics, Yale University, New Haven, CT 06511}

\begin{abstract}
Understanding the detailed structure of energy flow within jets, a field known as jet substructure, plays a central role in searches for new physics, and precision studies of QCD. Many applications of jet substructure require an understanding of jets initiated by heavy quarks, whose description has lagged behind remarkable recent progress for massless jets.
In this \emph{Letter}, we initiate a study of correlation functions of energy flow operators on beauty and charm jets to illuminate the effects of the intrinsic mass of the elementary particles of QCD. 
We present a factorization theorem incorporating the mass of heavy quarks, and show that the heavy quark jet functions for energy correlators have a simple structure in perturbation theory. Our results achieve the very first full next-to-leading-logarithmic calculation of the heavy quark jet substructure observable at the LHC.
Using this framework, we study the behavior of the correlators, and show that they exhibit a clear transition from a massless scaling regime, at precisely the scale of the heavy quark mass. 
This manifests the long-sought-after dead-cone effect and illustrates fundamental effects from the intrinsic mass of beauty and charm quarks in a perturbative regime, before they are confined inside hadrons. 
Our theoretical framework for studying energy correlators using heavy jets has many exciting applications for improving the description of mass effects in next generation parton shower event generators, probing the QGP, and studying heavy flavor fragmentation functions. 
\end{abstract}

\maketitle

\emph{Introduction.}---Jet substructure is playing an increasingly central role in collider physics  \cite{Larkoski:2017jix,Kogler:2018hem,Marzani:2019hun}, with applications ranging from innovative searches for new physics, to unravelling the nature of the quark-gluon plasma (QGP) \cite{Connors:2017ptx,Busza:2018rrf,Andrews:2018jcm,Cunqueiro:2021wls,Apolinario:2022vzg}. The progressively subtle features of the radiation pattern being exploited have motivated significant progress in the description of the perturbative substructure of jets, including the introduction of new theoretically motivated observables \cite{Chen:2020vvp,Chen:2022swd} and new theoretical techniques \cite{Hofman:2008ar,Kologlu:2019mfz,Chen:2021gdk,Chen:2022jhb,Chang:2022ryc}.

Despite this tremendous progress, many physical applications require an understanding of the substructure of jets initiated by massive (beauty or charm) jets. The most famous examples in the context of beyond the Standard Model searches, being the search for $H\to b \bar b$, which initiated the field of jet substructure \cite{Butterworth:2008iy}, and recent searches for $H\to c \bar c$, for which the strongest constraints come from jet substructure \cite{Qu:2019gqs,CMS:2022jed,CMS:2022pnq}. In the context of QCD, heavy quarks provide direct access to mass effects of quarks before their confinement into hadrons, and provide interesting probes of the QGP \cite{Armesto:2003jh,Dokshitzer:2001zm,PHENIX:2006iih,Moore:2004tg,Andronic:2015wma}. For a wealth of other discussions of the importance of heavy quarks, see e.g. studies of heavy quark production \cite{Czakon:2021ohs,Czakon:2011xx,Czakon:2013goa,Czakon:2012pz,Barnreuther:2012wtj,Czakon:2012zr,Czakon:2007ej,Czakon:2008ii,Mitov:2006xs,Catani:2020kkl,Nason:1989zy,Nason:1987xz,Mangano:1991jk,Frixione:1994nb,Frixione:1997ma,Cacciari:2012ny,Cacciari:2005rk,Nason:1989zy,Ellis:1988sb,Kang:2016ofv,Chen:2017qda,Chen:2022vzo}, fragmentation \cite{Mele:1990cw,Mele:1990yq,Melnikov:2004bm,Mitov:2005vk,Mitov:2004du,Melnikov:2004bm,Fickinger:2016rfd}, massive event shapes~\cite{Lepenik:2019jjk,Bris:2020uyb,L3:2004cdh,L3:2008lqk}, jet substructure \cite{Bain:2016clc,Mehtar-Tani:2016aco,Dehnadi:2016snl,Kang:2017frl,Casalderrey-Solana:2019ubu,Andrews:2018jcm,Lee:2019lge,Chien:2010kc,Makris:2018npl,Li:2018xuv,Dai:2018ywt} and experimental measurements \cite{SLD:1996yvs,SLD:2003qfy,DELPHI:2005zht,ALICE:2019cbr,CMS:2020geg,ATLAS:2021agf,ALICE:2020fyz}.

The theoretical description of the substructure of heavy quark jets is made difficult by the presence of the additional scale, and has seen much less progress than its massless counterpart. This is true both for perturbative calculations, as well as for parton shower simulations where large discrepancies with data are often observed  \cite{ALICE:2019cbr,CMS:2020geg,ATLAS:2021agf,ALICE:2020fyz}. While many beautiful measurements of event shapes on b-quark jets were made at SLD/LEP \cite{SLD:1996yvs,SLD:2003qfy,DELPHI:2005zht}, a particular complexity associated with the high energies of the LHC is that it requires a description of the transition from a regime where the quark behaves as massless to one where it behaves as massive. Motivated both by the wealth of physical applications, and by the development of a next generation of parton showers \cite{Li:2016yez,Hoche:2017hno,Hoche:2017iem,Dulat:2018vuy,Gellersen:2021eci,Hamilton:2020rcu,Dasgupta:2020fwr,Hamilton:2021dyz,Karlberg:2021kwr} that must incorporate heavy quark effects, we believe that the study of heavy quark jets requires renewed attention and a new approach.

\begin{figure}
\includegraphics[scale=0.36]{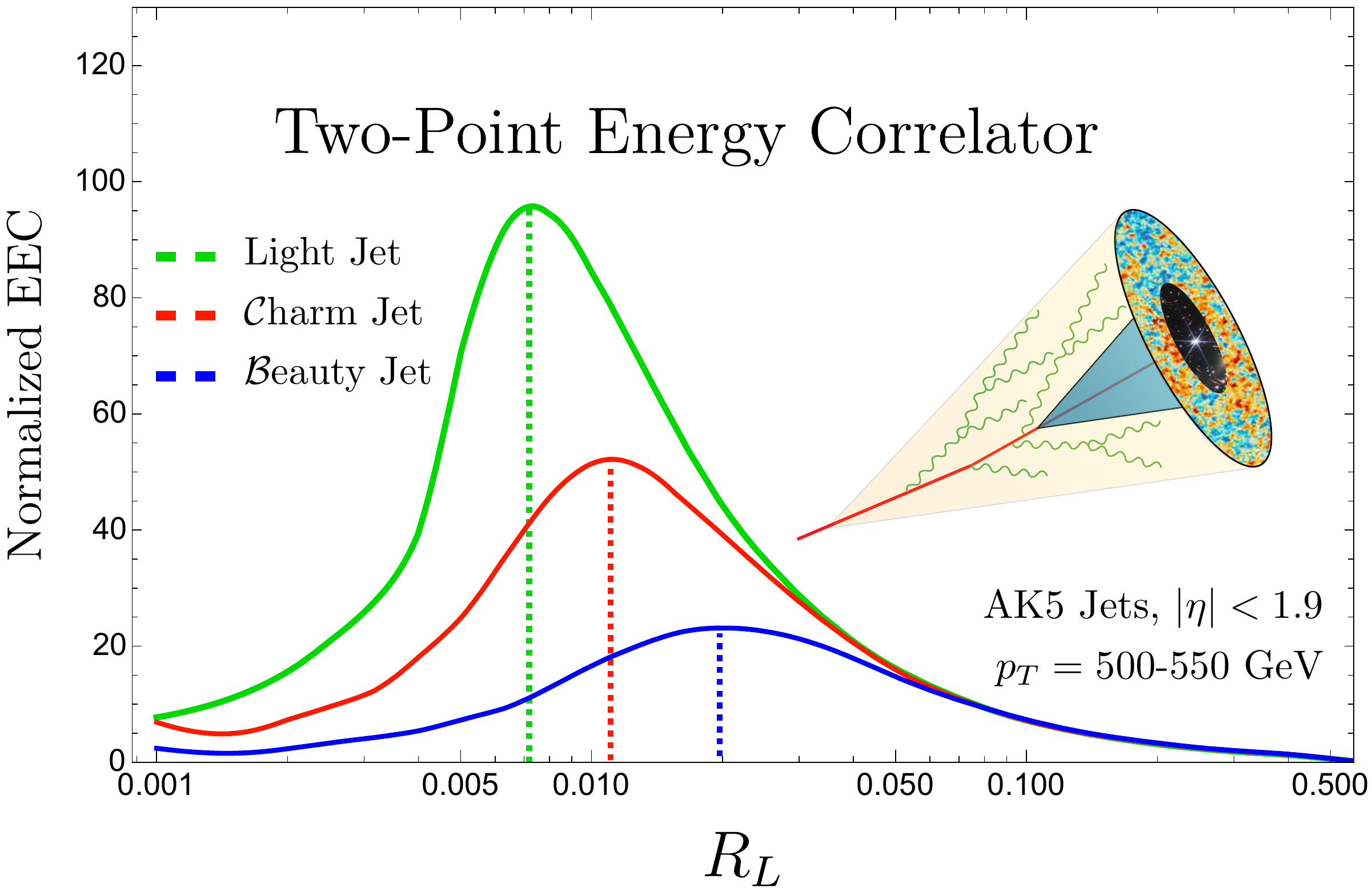}
\caption{The EEC measured inside high-$p_T$ jets. The radiation pattern in the jet is modified by the presence of heavy quarks masses, suppressing small angle radiation and giving rise to a ``dead cone". }
\label{fig:jet}
\end{figure} 

In this \emph{Letter} we initiate a study of heavy quark jets using correlation functions \cite{Basham:1979gh,Basham:1978zq,Basham:1978bw,Basham:1977iq} $\langle \Psi | \mathcal{E}(\vec n_1) \mathcal{E}(\vec n_2) \cdots \mathcal{E}(\vec n_k)  |\Psi \rangle$ of energy flow operators \cite{Belitsky:2001ij,Korchemsky:1999kt,Korchemsky:1997sy,Hofman:2008ar}, $\mathcal{E}(\vec n)$, allowing us to apply recent theoretical progress in this area\footnote{For a selection, see e.g. \cite{Hofman:2008ar,Belitsky:2013xxa,Belitsky:2013bja,Belitsky:2013ofa,Korchemsky:2015ssa,Belitsky:2014zha,Dixon:2018qgp,Luo:2019nig,Henn:2019gkr,Chen:2019bpb,Dixon:2019uzg,Korchemsky:2019nzm,Chicherin:2020azt,Kravchuk:2018htv,Kologlu:2019bco,Kologlu:2019mfz,Chang:2020qpj,Dixon:2019uzg,Chen:2020uvt,Chen:2020vvp,Chen:2019bpb,Chen:2020adz,Chicherin:2020azt,Chen:2021gdk,Korchemsky:2021okt,Korchemsky:2021htm,Chang:2022ryc,Chen:2022jhb,Yan:2022cye,Yang:2022tgm} and in particular the development of the light-ray operator product expansion (OPE) \cite{Hofman:2008ar,Kologlu:2019mfz,Chang:2020qpj} and celestial blocks \cite{Kologlu:2019mfz,Chang:2020qpj,Chen:2022jhb,Chang:2022ryc}. } to study the effects of intrinsic quark masses within jets. We present a factorization formula describing the universal behavior of the correlators in the small angle limit, and compute perturbatively the effects of the heavy quark mass, showing that they take a simple structure. We show that the heavy quark mass imprints itself as a characteristic angular scale in the energy correlator, see \Fig{fig:jet}, allowing us direct access to the ``dead-cone" effect \cite{Dokshitzer:1991fd} recently measured by the ALICE collaboration \cite{Cunqueiro:2018jbh,Zardoshti:2020cwl,ALICE:2021aqk,alicenature}, but in an observable that can be systematically computed in perturbation theory. This extends the list of physical systems to which the energy correlators have recently been applied to derive new insights, ranging from top quark decays \cite{Holguin:2022epo}, to nucleon structure \cite{Liu:2022wop}, and massless jets in the QGP \cite{Andres:2022ovj}.


 \emph{Factorization Theorem.}---Since our interest is on the behavior of energy correlators \emph{inside} jets, we are naturally led to study their universal behavior at small angles, which can be described by universal factorization theorems. Here we present the factorization theorem for a general class of scaling observables derived from the $N$-point correlation functions by integrating out all the information about the shape while keeping the length of the longest side, $R_L$, fixed, which are referred to as projected correlators, and will be denoted ENC \cite{Chen:2020vvp}. 
In the small angle limit, the cumulant of the projected $N$-point correlators $\Sigma^{[N]}(R_L,p_T^2, m_Q,\mu)$ factorizes into a hard function $\vec H(x,p_T^2,\mu)$, which describes the production of the collinear source, and the energy correlator jet function, $\vec J^{[N]}(R_L,x,\mu,m_Q)$, which describes the $R_L$ dependence of the observable
 \begin{align}\label{eq:factorizationformula}
 &\Sigma^{[N]}(R_L,p_T^2, m_Q,\mu) \\
 &\hspace{1cm}=\int_{0}^{1} dx\, x^N\, \vec J^{[N]} (R_L,x,m_Q,\mu) \cdot \vec H (x,p_T^2,\mu)\,. \nn
 \end{align}
This factorization was first derived in the context of $e^+e^-$~\cite{Dixon:2019uzg} and was recently extended to hadron colliders~\cite{Lee:2022ige}. It should be viewed as an extension of the classic factorization theorems  \cite{Collins:1981ta,Bodwin:1984hc,Collins:1985ue,Collins:1988ig,Collins:1989gx,Collins:2011zzd,Nayak:2005rt} for massive quark fragmentation \cite{Mele:1990cw,Mele:1990yq,Melnikov:2004bm,Mitov:2005vk,Mitov:2004du,Melnikov:2004bm,Fickinger:2016rfd} to the case of jet substructure observables. Note that massive jets may introduce additional subtleties~\cite{Banfi:2006hf,Banfi:2007gu,Gauld:2022lem,Caletti:2022glq} in the factorization theorem beyond the next-to-leading-logarithmic (NLL) accuracy we work in this paper. In the case of hadron colliders, $\vec H(x,p_T^2,\mu)$ also includes the parton distribution functions of the incoming beams as well as the matching coefficients incorporating the details of the jet algorithm. The hard function in proton-proton was recently computed at NNLO \cite{Czakon:2021ohs}, which should allow us to extend the perturbative accuracy of our calculation. The corresponding hard scale of the hadron colliders is associated with the transverse momentum of the jet, $\mu_H\sim p_T$. When jet algorithms such as the anti-$k_T$ jet algorithm~\cite{Cacciari:2005hq,Salam:2007xv,Cacciari:2008gp,Cacciari:2011ma} are used, they also introduce a jet scale associated with the jet radius as $\mu_J \sim p_T R$. Compared to what is presented in literature, we now also include the heavy quark mass in our factorization framework. See also~\cite{Nayak:2005rt,Collins:1989gx}. As long as $\mu_J \gg m_Q$, the heavy quark mass dependence factorizes completely into the heavy flavor jet function. That is, the hard function and the jet function are vectors in flavor space such that $\vec H \equiv \lbrace H_{g}, H_{q}, H_{Q} = H_{q} \rbrace$ and $\vec J^{[N]}\equiv \lbrace J^{[N]}_{g}, J^{[N]}_{q}, J^{[N]}_{Q}\rbrace$, where $g, q$ and $Q$ stand for gluon, light quark and massive quark labels respectively. 

Since factorization theorems involving massive quarks are much less well tested than their massless counterparts, to verify the factorization theorem in Eq.~\ref{eq:factorizationformula}, we have computed the full angle EEC for massive quarks in $e^+e^-\to Q \bar Q g$, expanded it in the small angle limit, and used it to separately extract the heavy quark EEC jet function.  We find exact agreement with the direct calculation of the jet function from its operator definition in the factorization theorem, which is presented in Eq.~\ref{eq:oneloopjet}, providing a highly non-trivial check on the factorization theorem. This also highlights the perturbative tractability of the energy correlators on massive partons, also highlighted by the perturbative calculations beyond the leading order for massless partons \cite{Belitsky:2013ofa,Dixon:2018qgp,Luo:2019nig,Henn:2019gkr}, and is deserving of further study.

\begin{figure}
\includegraphics[scale=0.27]{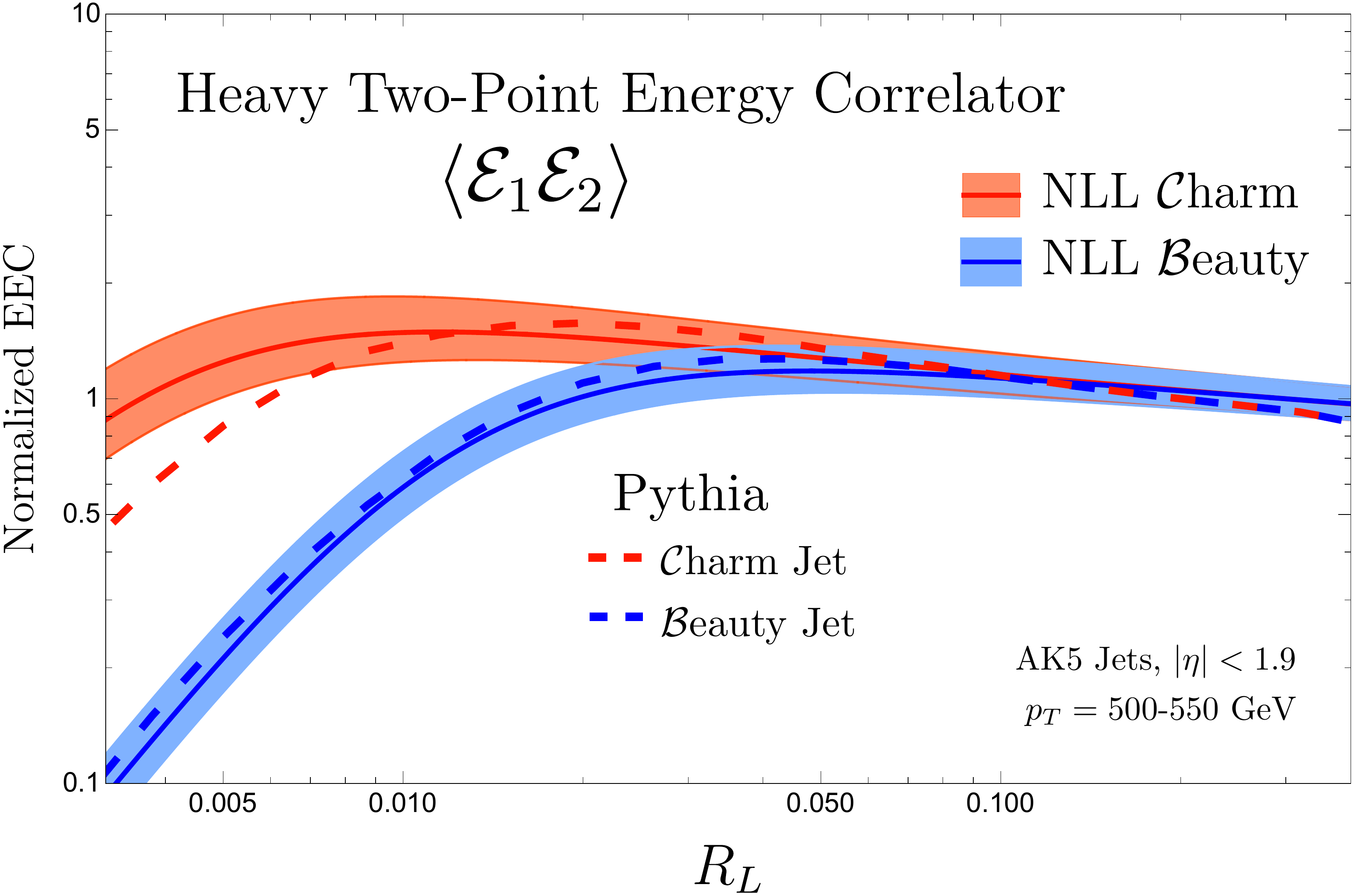} 
\caption{The EEC for beauty and charm jets illustrating a UV scaling behavior at large angles, and a mass dependent suppression at small angles.}
\label{fig:twopointEEC}
\end{figure}

\emph{The Heavy Quark EEC Jet Function.}---The cumulant $N$-point heavy quark energy correlator jet function is defined as \cite{Dixon:2019uzg}
\begin{align}\label{eq:jetdef}
J^{[N]}_{Q}(R_L,m_Q)=&\sum_{X} \sum_{i_1, i_2,...,i_N \in X}\left\langle 0\left|\bar{\chi}_{n}\right| X\right\rangle \frac{E_{i_1} E_{i_2}\cdots E_{i_N}}{p_T^N}\nonumber\\
& \Theta\left(\text{max}\{\theta_{i j}\}<R_L\right)\left\langle X\left|\chi_{n}\right| 0\right\rangle\,,
\end{align}
where $\chi_{n}$ is the gauge invariant collinear field in SCET for the massive quark field~\cite{Bauer:2000ew, Bauer:2000yr, Bauer:2001ct, Bauer:2001yt, Bauer:2002nz}. We choose to retain only the maximum angle information here, which corresponds to the $N$-point projected energy correlator discussed below. 
As the UV physics is independent of the heavy quark mass, the UV-poles of the heavy quark jet functions are identical to those of the light quark jet functions.

The light quark energy correlator jet function was computed in \cite{Dixon:2019uzg} by integrating the Altarelli-Parisi splitting functions \cite{Altarelli:1977zs}  over the massless collinear phase space. In the case of a massive quark, the phase space and the matrix elements are nontrivially modified due to the presence of the mass. Representative one-loop diagrams are
\begin{align}\label{eq:jet_diagrams}
\ \raisebox{0.0cm}{\fd{2.15cm}{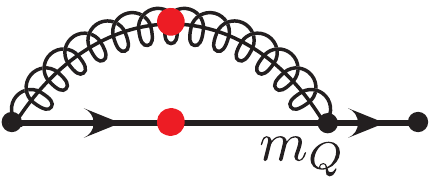}}\ +\ \raisebox{0.0cm}{\fd{2.15cm}{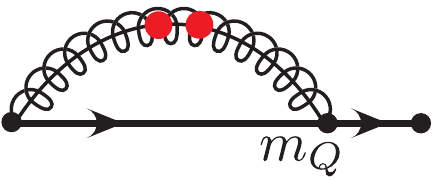}}+\ \raisebox{0.0cm}{\fd{2.15cm}{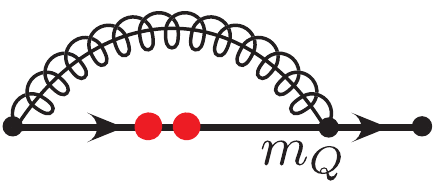}}\,. \nn
\end{align}
We have computed the jet function both using the massive splitting functions \cite{Catani:2000ef}, as well as massive \cite{Leibovich:2003jd} SCET \cite{Bauer:2000ew, Bauer:2000yr, Bauer:2001ct, Bauer:2001yt, Bauer:2002nz}, finding agreement.

\begin{figure}
\includegraphics[scale=0.33]{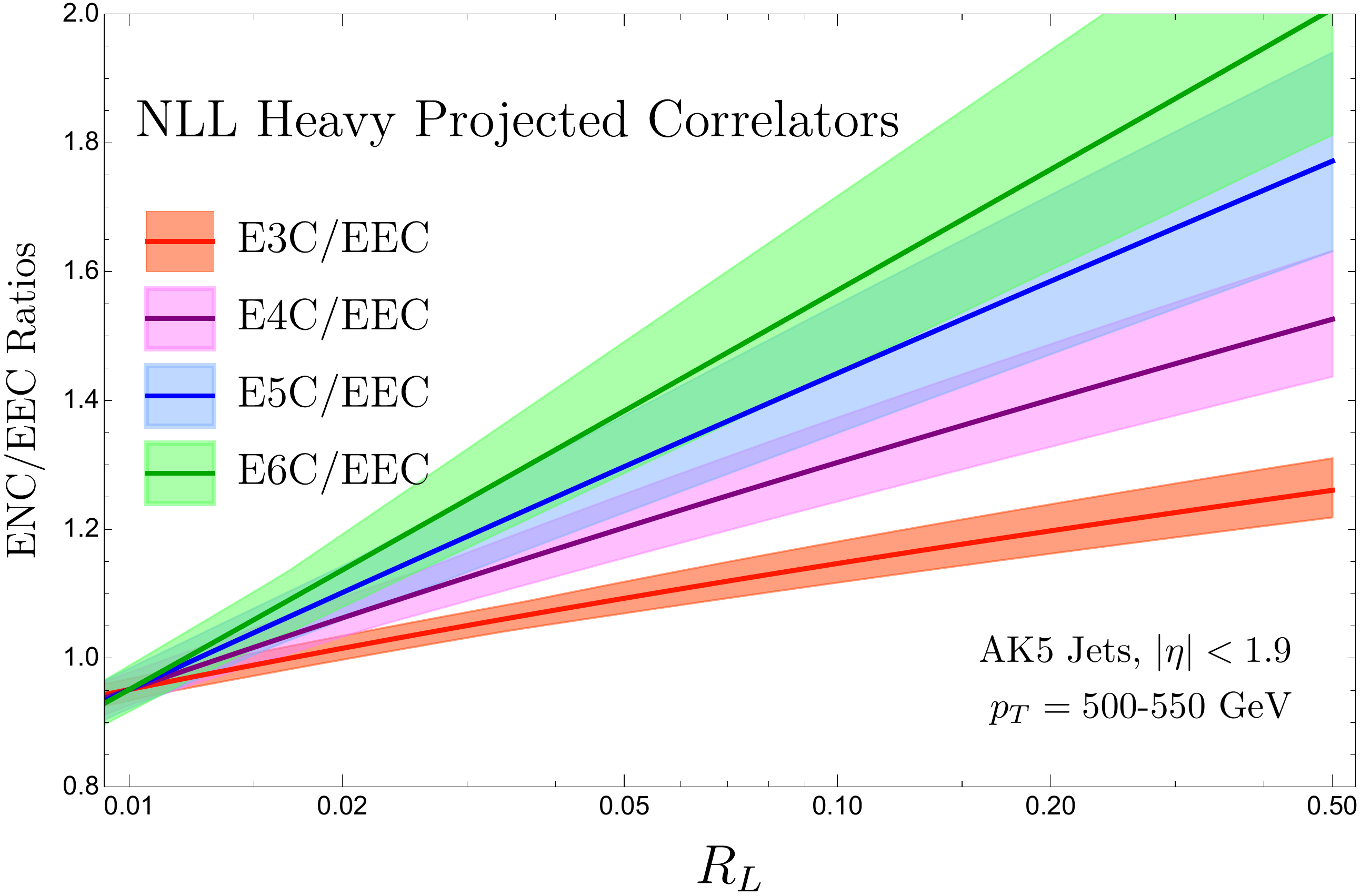} 
\caption{Ratios of the multi-point projected correlators resolve the structure of the UV scaling dimensions of QCD, and are independent of the heavy quark masses.} 
\label{fig:ratioplot_a}
\end{figure}

An interesting feature of the massive jet function calculation is that the heavy quark mass renders the virtual diagrams, as well as diagrams where both detectors lie on the same particle, not scaleless, and sensitive to the quark mass.
 With the mass regulating the IR-poles, the dimensional regularization parameter $\epsilon$ only regulates the UV-poles, and there is no ambiguity with respect to the nature of different poles. The IR sensitivity in the poles cancels in the sum of the diagrams as expected and the remaining UV divergence corresponds to the LO twist-2 spin-N+1 anomalous dimension as in the massless case \cite{Dixon:2019uzg}. This consistency is also required for the consistency of the factorization theorem given in Eq.~\eqref{eq:factorizationformula}. Suppressing the contact terms for simplicity, for $R_L\neq 0$, the NLO heavy quark jet function is given by
\begin{widetext}
\begin{equation}
\begin{aligned}
J^{[N]}_{Q}(R_L,m_Q)|_{R_L\neq 0} =& \frac{\alpha_s C_F}{4\pi}\int dx \frac{2(1-(1-x)^N-x^N)\left[2x^3+(1+x^2)(x+\delta)(x+\bar{\delta})\ln\frac{\delta \bar{\delta}}{(x+\delta)(x+\bar{\delta})}\right]}{(-1+x)(x+\delta)(x+\bar{\delta})}\,,
\end{aligned}
\label{eq:oneloopjet}
\end{equation}
\end{widetext}
where $\delta = \frac{i m_Q}{p_T R_L}$. This integral can be done as a function of $N$ in terms of hypergeometric functions, however, the particular result is not particularly enlightening. For integer $N$, the results are expressed in terms of the transcendental functions of weight 1 (i.e. logarithms) with alphabet $\{\delta,\bar{\delta}, 1+\delta, 1+\bar{\delta}\}$, and explicit results for up to $N=6$, as well as analogous results for the mass-dependent part of the gluon jet function, are presented in the \emph{Supplemental Material}. It would be interesting to understand if this simple structure for the heavy quark jet function persists at higher loop order.


\emph{The Two-Point Energy Correlator.}---The lowest order correlation function that exhibits a non-trivial dependence on the angular distance is the two-point energy correlator. This corresponds to taking $N=2$ in Eq.~\eqref{eq:factorizationformula}. As the angular scale $R_L$ is associated with the transverse momentum exchange $\sim p_T R_L + m_Q$ between the two-point, we expect to observe distinct regimes as the angular scale is varied. In Fig.~\ref{fig:twopointEEC}, we compare our analytic calculation of the two-point energy correlators for beauty and charm quark jets with results from the parton shower Monte Carlo Pythia 8.2~\cite{Sjostrand:2014zea}.\footnote{Note that this figure contains identical information as the Fig.~\ref{fig:jet}, but is plotted as log-log scales to emphasize the scaling behaviors.} We have verified that Vincia~\cite{Giele:2007di,Gehrmann-DeRidder:2011gkt,Fischer:2016vfv} gives compatible results. In Pythia we select $gg\to Q\bar{Q}$ and $q\bar{q} \to Q\bar{Q}$ processes, where $Q$ is beauty or charm or light flavor quarks. In order to carry out a corresponding theoretical calculation, we choose $H_q = H_g = 0$ in Eq.~\eqref{eq:factorizationformula}, whereas this may experimentally correspond to using experimental tagging techniques to select the heavy flavor jets~\cite{ALICE:2019cbr,CMS:2020geg,ATLAS:2021agf,ALICE:2020fyz}. Our analytic calculation is performed at NLL (single log counting), using the ingredients developed in \cite{Lee:2022ige}, including the NLO hard functions \cite{Aversa:1988mm,Aversa:1988fv,Aversa:1988vb,Aversa:1989xw,Aversa:1990uv,Jager:2004jh}, the NLO fragmenting jet functions \cite{Kang:2016mcy,Kang:2016ehg} and the NLO energy correlator jet constants \cite{Dixon:2019uzg}. This is the first calculation of a heavy quark jet substructure observable at the LHC at this perturbative order.

At scales larger than the heavy quark masses, we observe that the correlators exhibit a scaling behavior identical to that for massless quarks (We will verify this quantitatively in the next section). As  $ R_L \to m_Q/p_T$, we observe an onset of sensitivity to the heavy quark mass scale. A similar behavior was observed for light quark jets, but with the turn over at a scale of $ R_L \to \Lambda_{\text{QCD}}/p_T$ \cite{Komiske:2022enw,Liu:2022wop}. An interesting feature of heavy quark jets is that this turn over is described by perturbation theory, in particular, the heavy quark EEC jet function in Eq.~\ref{eq:oneloopjet}. Using a profile function to match between the resummation in the scaling and the heavy quark mass regions, we find good agreement with parton shower simulations, particularly for $b$-quark jets, where the turn over is more in the perturbative region. We believe this turn over region is particularly interesting for improving the description of heavy quark dynamics in parton shower programs. To our knowledge, ours is the first infrared safe heavy flavor observable (only sensitive to collinear dynamics) that exhibits such strong sensitivity to the intrinsic mass effects. We will now study the UV (large angle) and IR (small angle) regimes in more detail.

\begin{figure}
\includegraphics[scale=0.33]{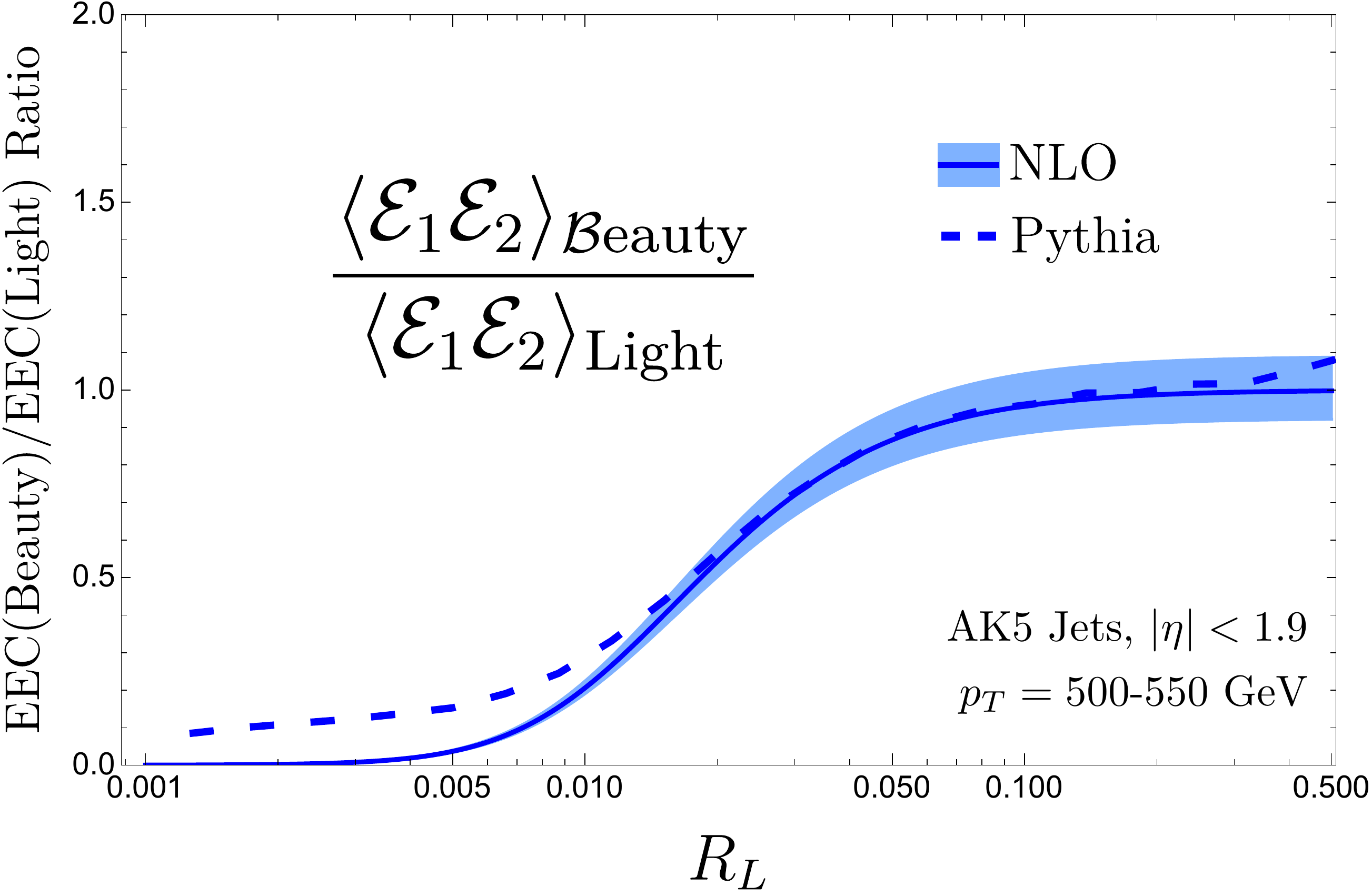} 
\caption{Ratios of heavy/light correlators reveal a suppression of the small-angle gluon radiation at the angular scale of the heavy quark mass. This provides a calculable observable sensitive to the dead-cone effect.} 
\label{fig:ratioplot_b}
\end{figure}

\emph{Projected Correlators and Anomalous Dimensions.}---A particularly interesting feature of the high LHC energies is that we are able to clearly see the transition between a regime where the heavy quark behaves as massless to a regime where the mass dominates. To make this quantitative, we can study ratio of projected correlators. This corresponds to taking the ratio of $N>2$ and $N=2$ point correlators calculated using Eq.~\eqref{eq:factorizationformula}. These ratios were designed to remove IR effects and isolate UV scaling associated with twist-2 anomalous dimensions \cite{Chen:2020vvp}. Although this was originally intended to eliminate non-perturbative effects at the scale $\Lambda_{\text{QCD}}$, we find that it also cancels the heavy quark effects, which are also IR. In Fig.~\ref{fig:ratioplot_a} we show the projected correlator ratios up to six-points for massive quark, which exhibit a clear scaling behavior, and is identical to the massless calculations presented in \cite{Lee:2022ige}.\footnote{The difference compared to \cite{Lee:2022ige} only arises from making the choice $H_q = H_g = 0$ in Eq.~\eqref{eq:factorizationformula} as discussed, and ratios are otherwise identical regardless of whether one measures them on a charm or beauty or massless jet.} We find identical scaling behavior to the massless case. While this is expected since the anomalous dimensions are a UV effect, we find it remarkable that we can isolate them in a clean manner.

\emph{The Dead-Cone Effect.}---In the other extreme, we can isolate infrared effects by taking the ratio of the two-point correlator on massive and massless jets. Since the scaling behavior is the same in both cases, this ratio should be calculable in fixed order perturbation theory, and be reflective of the infrared dynamics of the quark mass. This ratio is shown in Fig.~\ref{fig:ratioplot_b}, where we find excellent agreement between our fixed order calculation and the Pythia parton shower, illustrating that the transition region is under perturbative control.

This shape of the ratio in Fig.~\ref{fig:ratioplot_b} is a clean manifestation of what is commonly referred to as the ``dead-cone" effect, which is a fundamental feature of all gauge field theories and is a direct signature of the intrinsic mass before confinement~\cite{Dokshitzer:1991fd}. The dead-cone effect was recently observed experimentally \cite{alicenature} using an observable based on a de-clustering procedure \cite{Cunqueiro:2018jbh}. Here we have shown how it can be seen in the energy-correlators, where the description of the transition region can be computed systematically in perturbation theory. We look forward to real data observation of the dead-cone effect using energy correlators, and comparison to higher order calculations of heavy quarks in the near future.


 \emph{Conclusions.}---In this \emph{Letter} we have initiated a study of heavy quark jets using energy correlators. While the study of heavy quarks has long played an important role in QCD for understanding the mass effects of elementary particles, much less attention has been devoted to their study in jet substructure, primarily due to the prohibitive complexity associated with the additional mass scale.
 
We derived a factorization theorem for projected energy correlators on heavy quark jets, and showed that the jet functions, which encode the heavy quark mass effects can be computed analytically, and take a simple form.  We demonstrated that the scaling of heavy energy correlators, like that of light energy correlators, is also governed by the twist-2 anomalous dimensions at large angles, but shows a turnover at angles of order the heavy quark mass. This provides an observable that is simultaneously experimentally measurable, and can be computed systematically in perturbation theory, that manifests the long-sought-after dead-cone effect~\cite{Dokshitzer:1991fd,Zardoshti:2020cwl,ALICE:2021aqk,alicenature,Armesto:2003jh,Thomas:2004ie,Maltoni:2016ays,Cunqueiro:2018jbh}.

We believe that a precision understanding of the radiation pattern of jets initiated by massive quarks, as provided by the energy correlators, has many important applications. First, they can be used to improve the description of heavy quarks within next  generation parton showers \cite{Li:2016yez,Hoche:2017hno,Hoche:2017iem,Dulat:2018vuy,Gellersen:2021eci,Hamilton:2020rcu,Dasgupta:2020fwr,Hamilton:2021dyz,Karlberg:2021kwr}, which play a crucial role in many LHC searches, such as $H\to c\bar c$.  For this purpose it will also be important to extend our results to shape dependent multi-point correlators~\cite{Chen:2019bpb,Chen:2022swd}. Second, heavy quark jets are a powerful probe of the quark-gluon plasma, and our calculation provides a first step towards understanding how they are modified by interactions with the medium. Finally, it would be interesting to apply our approach to other heavy quark systems, such as quarkonia to help in understanding the fragmentation mechanisms of heavy flavor hadrons and in resolving outstanding puzzles of heavy quarkonium production~\cite{Andronic:2015wma,Nayak:2005rt,Fickinger:2016rfd,Zhang:2004qm,Mitov:2004du,Mitov:2005vk,Melnikov:2004bm,Mele:1990yq,Brambilla:2010cs,QuarkoniumWorkingGroup:2004kpm,Bodwin:1994jh,Bain:2017wvk,Kang:2017yde}. We hope that the beauty and charm of the energy correlators will open many new avenues to study heavy particles in QCD.

\section{Acknowledgements}
We thank Christine Aidala, Shanshan Cao, Paul Caucal, Stefan Hoche, Weiyao Ke, Kara Mattioli, Jared Reiten, Peter Skands, Alba Soto-Ontoso, Varun Vaidya, and Nima Zardoshti
for their helpful comments, discussions, and questions, as well as collaboration on related work. We thank Jack Holguin, Johannes Michel, and Jingjing Pan for comments on the draft. I.M. thanks Philip Burrows, Jim Brau and Michael Peskin for making him aware of the existence of SLD. K.L. was supported by the LDRD program of LBNL and the U.S. DOE under contract number DE-AC02-05CH11231 and DE-SC0011090. B.M. and I.M. are supported by start-up funds from Yale University.
\bibliography{bquark_ref.bib}{}
\bibliographystyle{apsrev4-1}

\begin{widetext}

\section*{Supplemental Material}
\label{sec:supplemental}


\subsection*{Heavy Quark Jet Function}
In Eq.~\ref{eq:oneloopjet}, we presented a general result for the projected $N$-point correlators as a single fold integral. As discussed in the main text, because we choose $H_q=H_g=0$, the component of the gluon jet function involving heavy quarks do not contribute to our result. For completeness, the NLO expression of the mass-dependent part of the gluon jet function is given by
\begin{align}
 J^{[N]}_{g\to Q\bar{Q}}(R_L,m_Q)|_{R_L\neq 0} =& \frac{\alpha_s T_F}{4\pi}\int dx \frac{2(1-(1-x)^N-x^N)\left[2(1-x)^3x^3+\left(1-2x(1-x)\right)\left(x^2(1-x)^2+\delta \bar{\delta}\right)\ln\frac{\delta \bar{\delta} + x^2(1-x)^2}{\delta\bar{\delta}}\right]}{\left(x^2(1-x)^2+\delta \bar{\delta}\right)}\,,
\end{align}
where $\delta = \frac{iM}{p_TR_L}$.

Here we collect explicit expressions for the projected correlators for the NLO heavy quark jet function in Eq.~\ref{eq:oneloopjet} and $g\to Q\bar{Q}$ component of the NLO gluon jet function up to $N=6$. When $R_L\neq 0$, they are given up to $N=6$ by
\bea
J_Q^{[2]}|_{R_L\neq 0}=&\frac{\alpha_s C_F}{4\pi}\left\{\left[\delta^4-4\delta^3+2\delta^2-3\right]\ln\left(\frac{\delta}{1+\delta}\right)-\frac{1}{2}\left(9\delta^2+\frac{31}{6}\right)\right\}+c.c\,,\\
J_Q^{[3]}|_{R_L\neq 0}=&\frac{\alpha_s C_F}{4\pi}\left\{\left[\frac{3}{2}\delta^4-6\delta^3+3\delta^2-\frac{9}{2}\right]\ln\left(\frac{\delta}{1+\delta}\right)-\frac{1}{2}\left(\frac{27}{2}\delta^2+\frac{31}{4}\right)\right\}+c.c\,,\\
J_Q^{[4]}|_{R_L\neq 0}=&\frac{\alpha_s C_F}{4\pi}\left\{\left[\frac{2}{3}\delta^6-\frac{16}{5}\delta^5-\delta^4-\frac{20}{3}\delta^3+4\delta^2-\frac{83}{15}\right]\ln\left(\frac{\delta}{1+\delta}\right)-\frac{1}{2}\left(\frac{106}{15}\delta^4+\frac{74}{5}\delta^2+\frac{1417}{150}\right)\right\}+c.c\,,\\
J_Q^{[5]}|_{R_L\neq 0}=&\frac{\alpha_s C_F}{4\pi}\left\{\left[\frac{5}{3}\delta^6-8\delta^5-5\delta^4-\frac{20}{3}\delta^3+5\delta^2-\frac{19}{3}\right]\ln\left(\frac{\delta}{1+\delta}\right)-\frac{1}{2}\left(\frac{53}{3}\delta^4+\frac{29}{2}\delta^2+\frac{107}{10}\right)\right\}+c.c\,,\\
J_Q^{[6]}|_{R_L\neq 0}=&\frac{\alpha_s C_F}{4\pi}\left\{\left[\frac{1}{2}\delta^8-\frac{20}{7}\delta^7-\frac{11}{3}\delta^6-\frac{84}{5}\delta^5-\frac{19}{2}\delta^4-6\delta^3+6\delta^2-\frac{734}{105}\right]\ln\left(\frac{\delta}{1+\delta}\right)\right.\nonumber\\
&\left.-\frac{1}{2}\left(\frac{87}{14}\delta^6+\frac{13477}{420}\delta^4+\frac{2767}{210}\delta^2+\frac{1030979}{88200}\right)\right\}+h.c\,,\\
J_{g\to Q\bar{Q}}^{[2]}|_{R_L\neq 0}=&\frac{\alpha_s T_F}{4\pi}\left\{\left[\frac{72\delta^3-24\delta^2-4\delta+2}{5\sqrt{1-4\delta}}\right]\ln\left(\frac{A-1}{A}\right)-\frac{1}{2}\left(-\frac{72}{5}\delta^2+\frac{28}{25}\right)\right\}+c.c\,,\\
J_{g\to Q\bar{Q}}^{[3]}|_{R_L\neq 0}=&\frac{\alpha_s T_F}{4\pi}\left\{\left[\frac{108\delta^3-36\delta^2-6\delta+3}{5\sqrt{1-4\delta}}\right]\ln\left(\frac{A-1}{A}\right)-\frac{1}{2}\left(-\frac{108}{5}\delta^2+\frac{42}{25}\right)\right\}+c.c\,,\\
J_{g\to Q\bar{Q}}^{[4]}|_{R_L\neq 0}=&\frac{\alpha_s T_F}{4\pi}\left\{\left[\frac{-1320\delta^4+3336\delta^3-992\delta^2-152\delta+76}{105\sqrt{1-4\delta}}\right]\ln\left(\frac{A-1}{A}\right)-\frac{1}{2}\left(-\frac{3116}{105}\delta^2+\frac{22634}{11025}\right)\right\}+c.c\,,\\
J_{g\to Q\bar{Q}}^{[5]}|_{R_L\neq 0}=&\frac{\alpha_s T_F}{4\pi}\left\{\left[\frac{-660\delta^4+912\delta^3-244\delta^2-34\delta+17}{21\sqrt{1-4\delta}}\right]\ln\left(\frac{A-1}{A}\right)-\frac{1}{2}\left(-\frac{802}{21}\delta^2+\frac{5143}{2205}\right)\right\}+c.c\,,\\
J_{g\to Q\bar{Q}}^{[6]}|_{R_L\neq 0}=&\frac{\alpha_s T_F}{4\pi}\left\{\left[\frac{728\delta^5-3700\delta^4+3560\delta^3-866\delta^2-110\delta+55}{63\sqrt{1-4\delta}}\right]\ln\left(\frac{A-1}{A}\right)-\frac{1}{2}\left(-\frac{104}{9}\delta^4-\frac{4946}{105}\delta^2+\frac{50696}{19845}\right)\right\}+c.c\,,
\eea
where $A = \frac{1}{2}(1-\sqrt{1-4\delta})$.

These were computed using the massive $1\to 2$ splitting functions, which we reproduce here for completeness, since they are less familiar than their massless counterparts. We have
\begin{align}
P_{Q\rightarrow Qg}(x,m_Q)= &\frac{C_F}{s-m_Q^2}\left[\frac{1+x^2}{1-x}-\epsilon(1-x)-\frac{2 m_Q^2}{s-m_Q^2}\right]\,,\\
P_{g\rightarrow  Q\bar Q}(x, m_Q)= &\frac{T_F}{s}\left[1-\frac{2 x(1-x)}{1-\epsilon}+\frac{2 m_Q^2}{(1-\epsilon) s }\right]\,,
\end{align}
where $s=(p_1+p_2)^2$, with $p_1$ and $p_2$ being the momentum of the two outgoing partons.

The virtual diagrams and the real diagrams with correlators on the same particle contribute to $R_L=0$ part of the heavy quark jet function. The IR sensitive poles cancel in the sum, and the remaining UV pole is regulated by dimensional regularization parameter $\epsilon$. They are given up to $N=6$ as 
\bea
J_Q^{[2]}|_{R_L= 0}=&\frac{\alpha_s C_F}{4\pi}\left\{-3\left[\frac{1}{\epsilon}+\ln\frac{\mu^2}{M^2}\right]-\frac{49}{6}\right\}\,,\\
J_Q^{[3]}|_{R_L= 0}=&\frac{\alpha_s C_F}{4\pi}\left\{-\frac{9}{2}\left[\frac{1}{\epsilon}+\ln\frac{\mu^2}{M^2}\right]-\frac{47}{4}\right\}\,,\\
J_Q^{[4]}|_{R_L= 0}=&\frac{\alpha_s C_F}{4\pi}\left\{-\frac{83}{15}\left[\frac{1}{\epsilon}+\ln\frac{\mu^2}{M^2}\right]-\frac{6611}{450}\right\}\,,\\
J_Q^{[5]}|_{R_L= 0}=&\frac{\alpha_s C_F}{4\pi}\left\{-\frac{19}{3}\left[\frac{1}{\epsilon}+\ln\frac{\mu^2}{M^2}\right]-\frac{779}{45}\right\}\,,\\
J_Q^{[6]}|_{R_L= 0}=&\frac{\alpha_s C_F}{4\pi}\left\{-\frac{734}{105}\left[\frac{1}{\epsilon}+\ln\frac{\mu^2}{M^2}\right]-\frac{193101}{9800}\right\}\,,\\
J_{g\to Q\bar{Q}}^{[2]}|_{R_L= 0}=&\frac{\alpha_s T_F}{4\pi}\left\{\frac{14}{15}\left[\frac{1}{\epsilon}+\ln\frac{\mu^2}{M^2}\right]\right\}\,,\\
J_{g\to Q\bar{Q}}^{[3]}|_{R_L= 0}=&\frac{\alpha_s T_F}{4\pi}\left\{\frac{11}{15}\left[\frac{1}{\epsilon}+\ln\frac{\mu^2}{M^2}\right]\right\}\,,\\
J_{g\to Q\bar{Q}}^{[4]}|_{R_L= 0}=&\frac{\alpha_s T_F}{4\pi}\left\{\frac{64}{105}\left[\frac{1}{\epsilon}+\ln\frac{\mu^2}{M^2}\right]\right\}\,,\\
J_{g\to Q\bar{Q}}^{[5]}|_{R_L= 0}=&\frac{\alpha_s T_F}{4\pi}\left\{\frac{11}{21}\left[\frac{1}{\epsilon}+\ln\frac{\mu^2}{M^2}\right]\right\}\,,\\
J_{g\to Q\bar{Q}}^{[6]}|_{R_L= 0}=&\frac{\alpha_s T_F}{4\pi}\left\{\frac{29}{63}\left[\frac{1}{\epsilon}+\ln\frac{\mu^2}{M^2}\right]\right\}\,.
\eea
As expected, the UV pole proportional to $\frac{1}{\epsilon}$ is given by the moment of the corresponding LO twist-$2$ spin-$N+1$ anomalous dimension. It is also worth nothing that in the limit $p_TR_L \gg M$, we recover the $1$-loop massless jet functions as expected.

\end{widetext}

\end{document}